\documentclass{openjournal}

\usepackage{amsmath,amssymb,mathrsfs,bm}

\DeclareMathAlphabet{\tens}{OT1}{cmss}{bx}{n}

\begin{document}

\title{Polarization cross-talk from non-stationary sources of image jitter}

\author{R.\ Casini \& A.\ G.\ de Wijn}

\affil{High Altitude Observatory,
NSF National Center for Atmospheric Research,\break
P.\ O.\ Box 3000, Boulder, CO 80307-3000, U.S.A.}

\begin{abstract}
We generalize a previous formulation for the estimation of polarization cross-talk induced by atmospheric seeing to the case of image jitter that must be described  
as a non-stationary, and generally non-isotropic, random process. We demonstrate how the newly derived formalism converges towards the previous one in the limit of stationary random image jitter. In the case of non-stationary processes, we compare the predictions of the two formalisms, and illustrate the salient differences. For the sake of demonstration, we adopt for our analysis a specific example of power-spectral density derived from a circulant auto-correlation function, which describes a class of stationary random processes periodic over the length of the time series, and  which we successively modified by the insertion of off-diagonal power spectral features at different frequencies in order to test the extended modeling framework for the estimation of polarization cross-talk.
\end{abstract}

\section{Introduction}

In \cite{Ca12} (hereafter, Paper~I; see also Erratum) we analyzed 
the role of atmospheric seeing in the
production of polarimetric cross-talk during a polarization modulation
sequence. That study allowed us to cast into a general formalism, and 
a set of physical assumptions, prior analyses of the seeing effect on 
polarization observations \citep{Li87,Ju04} that had used the power 
spectral density (PSD) profiles of sky scintillation data, specifically 
acquired to test the feasibility and efficiency of adaptive optic (AO)
systems, to investigate the benefit of AO correction with traditional
polarization modulation schemes. In particular, the analysis in Paper~I
rederived those previous results by interpreting the seeing displacement 
of the image at the detector as a \emph{stationary and ergodic stochastic 
process with zero mean amplitude, and isotropic over the detector plane}.

With this paper we want to extend such a formalism to the
case of image motion that is not stationary (hence, non-ergodic), 
and possibly non-isotropic (i.e., with different
spectral properties in the $x$ and $y$ directions). An example of such 
a motion is represented by some realizations of spacecraft jitter, 
when a pointing drift component is clearly detectable and uncompensated 
for over the time span of a polarimetric sequence. In such a case, the 
statistical properties (e.g., mean and variance) of the image motion 
become functions of the time of the observation, and the definition of 
a single-frequency (i.e., one-dimensional; 1D) PSD to fully characterize the image motion, analogously to the 
development of Paper~I, is frustrated by the lack of time-shift 
invariance of the auto-correlation function (ACF) of the motion's time series.
Instead, we must consider the spectral characteristics 
of the specific time-series realization of the jitter during a
polarimetric sequence, and the intrinsic two-time (i.e., bi-dimensional; 2D) nature of such an ACF naturally leads to the need to characterize the image motion via the corresponding 2D PSD\cite[see, e.g.,][]{Pa65}. 

Apart from these generalizations, the formalism and notation of Paper~I, and some of its hypotheses, remain unchanged. In particular, we retain the assumption that the jitter amplitude is small enough that its effects on the Stokes signals are linear in the displacement vector. This means that the estimation of any jitter-induced polarization cross-talk is still fundamentally driven by the presence of Stokes gradients on the image plane. Accordingly, we can retain the fundamental description of the instantaneous Stokes signal at a given point $\bm{x}_0$ on the image plane as given by Eq.~(8) of Paper~I. This is a legitimate assumption even for non-ergodic processes (whether stationary or not), as it only depends on the validity of the first-order expansion of the Stokes signals in the image displacement vector. The assumptions of isoplanatism of the field and of the constancy of the observed scene during the observation, which are also implied by the formalism of Paper~I (cf.~Eqs.~(1)-(7)), must of course also remain true.

While in the case of the atmospheric seeing considered in Paper~I stationarity was a consequence of the random phases realized during the evolution of the seeing phenomenon, in the present study non-stationarity arises from contributions to the image jitter that are not invariant under a time shift of the polarization modulation sequence, such as pointing drifts, mechanical and thermal transients (e.g., the activation of mechanisms on board; the transition of the spacecraft in or out of an eclipse), and periodic components (such has mechanical and electronic hums) that are phase-locked with the polarimetric sequence. In particular, periodic components typically stand out through the characteristic power spikes in the PSD of spacecraft jitter (REFS).\footnote{It is important to remark that the simple appearance of power spikes in a PSD is not sufficient to draw any conclusion on the phase coherence of their sources  and how these may correlate with a polarimetric sequence (see the concluding paragraph of Sect.~4).}
Correspondingly, the expectation values introduced in Sect.~2 must be understood as ensemble averages over independent realizations of the jitter time series defined relative to the same fixed initial time of the polarimetric sequence (see discussion at the end of Sect.~3), rather than as time averages over a single realization, as implied in Paper~I by the assumption of ergodicity of the seeing phenomenon.

Therefore, what changes in the current study is the interpretation of $\tilde{\bm{S}}(\bm{x}_0)$ and $\bm{\nabla}\tilde{\bm{S}}(\bm{x}_0)$ in Eq.~(8) of Paper~I, which no longer correspond to ensemble (long-time) averages over the realization of the image motion, but are directly the unperturbed Stokes vector and its gradient at the fixed position $\bm{x}_0$. As a consequence, the image motion is no longer required to have zero mean. Instead, the two-time ACF of Eq.~(\ref{eq:Gamma_def}) simply corresponds to the second-order moment of that displacement from $\bm{x}_0$, and therefore the formalism yields the mean-square error of the inferred Stokes parameters, which also includes the contribution of any systematic drift of the image.

In order to preserve as much as possible the notational framework of Paper~I, in the next section we will make the assumption that $\bm{x}_0=(0,0)$, without any loss of generality. With this convention, a useful refresher on Paper~I to follow the development of the next section will consist of the full reading of Sect.~3, the beginning of Sect.~5 up to and including Eq.~(27), and Sect.~6 up to the discussion of the numerical examples.

\section{Formulation of the problem}

As anticipated in the Introduction, we reference Paper~I for the basic notations and 
definitions. We then start from recalling the expression of the observed 
signal $\mathscr{I}_i$ for the modulation state $i=1,\ldots,n$ 
(see Eqs.~(9)-(11) of Paper~I):
\begin{eqnarray} \label{eq:signal_i}
\mathscr{I}_i
&=&\kappa\sum_{j=1}^4\,
      \int_{-\Delta t/2}^{+\Delta t/2}
      m_j(t+t_i)\,[S_j+\bm{\nabla} S_j\cdot\bm{x}(t+t_i)]\;\mathrm{d}t
	+\delta\mathscr{I}_i \nonumber \\
&=&\kappa\Delta t \sum_{j=1}^4 \biggl[
      m_{ij}\,S_j +\frac{1}{\Delta t} 
      \int_{-\Delta t/2}^{+\Delta t/2}
	m_j(t+t_i)\,\bm{\nabla} S_j\cdot \bm{x}(t+t_i)\;\mathrm{d}t
      \biggr] 
	+\delta\mathscr{I}_i \nonumber \\
&=&(\bar{\mathscr{I}}_i
	+\delta\mathscr{I}_i)
+\kappa \sum_{j=1}^4 
      \int_{-\Delta t/2}^{+\Delta t/2}
	m_j(t+t_i)\,\bm{\nabla} S_j\cdot \bm{x}(t+t_i)\;\mathrm{d}t\;,
\end{eqnarray}
where
\begin{equation} \label{eq:mod_mat}
m_{ij}\equiv\frac{1}{\Delta t} 
      \int_{-\Delta t/2}^{+\Delta t/2}m_j(t+t_i)\;\mathrm{d}t\;,
\end{equation}
are the elements of the ``average'' modulation matrix $\tens{M}$,
and $\kappa$ is a dimensional factor that accounts for the gain
of the detector during the exposure time $\Delta t$. 

The additional
error $\delta\mathscr{I}_i$, after subtraction of the dark signal, is a
random fluctuation with zero mean due to photon noise (see Paper~I,
Eq.~(9) and following discussion).
Because photon noise is statistically uncorrelated between different 
detector exposures, it only appears in the diagonal contributions to the 
polarimetric error $\sigma^2(S_i')$ (see Paper~I, Eq.~(15)). Therefore, it
will be dropped from the following derivation, as it can simply be
added back as a variance term (cf.~Paper~I, Eq.~(16) and following
discussion)
\begin{equation}
\sigma^2_{\rm ph}=\sum_{j=1}^n d_{ij}^2\,(\delta\mathscr{I}_j)^2\;,
\end{equation}
where $\tens{D}=(d_{ij})$ is the demodulation matrix, such that
$\tens{D}\tens{M}=\tens{1}$.\footnote{Generally, $\tens{D}$ is 
calculated as the Moore-Penrose inverse of $\tens{M}$, i.e.,
$\tens{D}=(\tens{M}^T \tens{M})^{-1}\tens{M}^T$, where $\cdot^T$ is
matrix transposition.}

We now must consider that the mean-square error on the observed 
Stokes parameter $S'_i$ produced over a polarimetric sequence of 
$N$ modulation cycles, $l=0,\ldots,N-1$, in the presence of a 
non-stationary (hence, non-ergodic) random image motion, cannot 
be expressed in terms of time-independent expectation values 
(cf.~Eq.~(15) of Paper~I), as the spectral characteristics of such 
a stochastic process will depend on the specific polarimetric 
sequence at hand, and in general on the particular modulation 
cycle of index $l$ in the sequence. 
Therefore, after dropping the photon-noise term,  we must rewrite 
Eq.~(\ref{eq:signal_i}) as
\begin{equation} \label{eq:signal_il}
\mathscr{I}_{i,l} \equiv \bar{\mathscr{I}}_i
	+\kappa \sum_{j=1}^4 
      \int_{-\Delta t/2}^{+\Delta t/2}
	m_j(t+t_{i,l})\,\bm{\nabla} S_j\cdot
\bm{x}(t+t_{i,l})\;\mathrm{d}t\;,
\qquad l=0,\ldots,N-1\;,
\end{equation}
where the first addendum corresponds to the  modulated signal that would be measured by the detector 
in the absence of image motion, whereas the second
addendum corresponds to the random error 
$\delta\mathscr{I}_{i,l}$ induced by image motion. 

The camera exposure interval $\Delta t$ is centered around the time 
\begin{equation} \label{eq:time}
t_{i,l}=(i-1)\frac{\Delta t}{r}+ l n \frac{\Delta t}{r} + \tau\;,
\end{equation}
where $r$ is the camera duty cycle ($0< r \le 1$) and $\tau$ is the
time offset between the midpoint of the first camera exposure and the
$0^\circ$ position of the modulator. For a stepped modulator, it is
always $\tau=0$, whereas for a continuously rotating
modulator, the two typical choices are $\tau=0$ ($0^\circ$ 
at midpoint of camera exposure) or $\tau=\Delta t/2$ ($0^\circ$ 
at beginning of camera exposure).

Our objective is to evaluate the (time dependent) expectation
value (cf.~Paper~I, Eq.~(15))
\begin{eqnarray} \label{eq:sigmaN}
\sigma^2_N(S'_i)\equiv E([S'_i-\bar{S}'_i]^2)
&=&\frac{1}{N^2}\sum_{j,k=1}^{nN} d_{ij}d_{ik}\,E\bigl(
	[\mathscr{I}_j-\bar{\mathscr{I}}_j]
	[\mathscr{I}_k-\bar{\mathscr{I}}_k]\bigr) \nonumber \\
&=&\frac{1}{N^2}\sum_{j,k=1}^n d_{ij}d_{ki}^T\sum_{l,l'=0}^{N-1} E\bigl(
	[\mathscr{I}_{j,l}-\bar{\mathscr{I}}_j]
	[\mathscr{I}_{k,l'}-\bar{\mathscr{I}}_k]\bigr) \nonumber \\
&=&(\tens{D}\,\tens{O}_n\tens{D}^T)_{ii}\;,
\end{eqnarray}
where, similarly to the case of $\bar{\mathscr{I}}_i$, the quantity $\bar{S}'_i$ corresponds to the true Stokes signal that would be observed in the absence of jitter, rather than to the expectation value $E(S_i')$ as in Paper~I.
The $n{\times}n$ error correlation matrix $\tens{O}_n$ is symmetric with elements
\begin{equation} \label{eq:O_jk}
O_{jk}
=\frac{1}{N^2}\sum_{l,l'=0}^{N-1}
	E([\mathscr{I}_{j,l}-\bar{\mathscr{I}}_j]
	  [\mathscr{I}_{k,l'}-\bar{\mathscr{I}}_k])\;.
\end{equation}
Its normalization by $1/N^2$ is needed so we can use the one-cycle demodulation 
matrix $\tens{D}=(d_{ij})$, to demodulate the $nN$ signals acquired over 
the full sequence of $N$ modulation cycles into the proper Stokes parameters (see Paper~I, Sect.~6).
The subscript $N$ in the mean-square error definition of Eq.~(\ref{eq:sigmaN}) stands to indicate 
explicitly that such an error estimate is specific to the 
sequence of $N$ modulation cycles under consideration.

We note that the matrix $\tens{O}_n=(O_{jk})$ is the expectation value
of the exterior (or Kronecker) product 
$\bm{\delta\mathscr{I}}\otimes\bm{\delta\mathscr{I}}^T$, where
$\bm{\delta\mathscr{I}}$ is the vector of components 
\begin{equation}
\delta\mathscr{I}_j
	\equiv
\frac{1}{N}\sum_{l=0}^{N-1}(\mathscr{I}_{j,l}-\bar{\mathscr{I}}_j)\;,
\end{equation}
averaged over the full sequence of $N$ modulation cycles.
Using Eq.~(\ref{eq:signal_il}) and standard analytic
manipulations adopted in Sect.~6 of Paper~I, we have
\begin{eqnarray} \label{eq:I_jk}
\delta\mathscr{I}_j
	\equiv
\frac{1}{N}\sum_{l=0}^{N-1}(\mathscr{I}_{j,l}-\bar{\mathscr{I}}_j)
&=& \frac{\kappa}{N}\sum_{p=1}^4
	\sum_{l=0}^{N-1}
	\int_{-\Delta t/2}^{+\Delta t/2}
	\textrm{d}t\;
	m_p(t+t_{j,0})\,
	\bm{\nabla}S_p\cdot\bm{x}(t+t_{j,l}) \nonumber \\
&=& \frac{\kappa}{N}\sum_{p=1}^4
	\sum_{l=0}^{N-1}
	\int_{-\infty}^{+\infty}
	\textrm{d}t\;
	\Pi_{\Delta t}(t)\,m_p(t+t_{j,0})\,
	\partial_\alpha S_p\,x_\alpha(t+t_{j,l})\;,
\end{eqnarray}
where we exploited the periodicity of $m_i(t)$ 
over a modulation cycle,
\begin{displaymath}
m_i(t+t_{j,l})=m_i(t+t_{j,0})\;,\qquad l=0,\ldots,N-1\;,
\end{displaymath}
and the implicit summation over $\alpha$ spans the two 
Cartesian coordinates of the image plane.
By introducing the functions
\begin{equation} \label{eq:Zcomp}
Z_{ij}(t)=\Pi_{\Delta t}(t)\,m_i(t+t_{j,0})\;,
\end{equation}
and using Eqs.~(\ref{eq:I_jk}) and (\ref{eq:Zcomp}) in Eq.~(\ref{eq:O_jk}),
we find at last (cf.~Paper~I, Eq.~(24))
\begin{equation} \label{eq:O_jk.expl}
O_{jk}= \frac{\kappa^2}{N^2}\sum_{p,q=1}^4
	\partial_\alpha S_p\,
	\partial_\beta S_q\,
	\sum_{l,l'=0}^{N-1}
	\int\!\!\!\int_{-\infty}^{+\infty}
	Z_{pj}(t) Z_{qk}(t')\,
	E(x_\alpha(t+t_{j,l})\,
	  x_\beta(t'+t_{k,l'}))\;
	\textrm{d}t\,
	\textrm{d}t'\;,
\end{equation}
where each of the indexes $\alpha$ and $\beta$ span the two
Cartesian coordinates of the image plane.

For a generally non-stationary random image motion, the 
expectation value in the
integral of Eq.~(\ref{eq:O_jk.expl}) corresponds to the
two-time ACF
\begin{equation} \label{eq:Gamma_def}
\Gamma_{\alpha\beta}(t,t')= E(x_\alpha(t)\, x_\beta(t'))\;.
\end{equation}
Similarly to the case of the ACF of a stationary
random process, we can introduce a PSD function associated to
$\Gamma_{\alpha\beta}(t,t')$ via 2D Fourier 
transform (see, e.g., \citealt{Pa65}, Eqs.~(12-22)-(12-23)),\footnote{In the analysis of non-stationary
processes leading to the definition of a two-time ACF,
it is customary to use opposite signs for the two frequencies entering
the 2D Fourier transform; see Eq.~(\ref{eq:2D-PSD}) below, and 
Eq.~(12-22) in \cite{Pa65}. Ultimately, this stems from the observation that, for 
stationary processes, the same opposite signs are produced when taking
the Fourier transform of the ACF $\Gamma(t)$ 
evaluated at the lag $t'-t$; see, e.g., Eq.~(38) of Paper~I.}
\begin{equation} \label{eq:2D-PSD}
P_{\alpha\beta}(\omega,\omega')=
	\int\!\!\!\int_{-\infty}^{+\infty}
	\Gamma_{\alpha\beta}(t,t')\,
	\mathrm{e}^{-\mathrm{i}(\omega t-\omega't')}\;
	\textrm{d}t\,
	\textrm{d}t'\;,
\end{equation}
with the inverse relation
\begin{equation} \label{eq:2D-PSD_inv}
\Gamma_{\alpha\beta}(t,t')=\frac{1}{(2\pi)^2}
	\int\!\!\!\int_{-\infty}^{+\infty}
	P_{\alpha\beta}(\omega,\omega')\,
	\mathrm{e}^{\mathrm{i}(\omega t-\omega't')}\;
	\textrm{d}\omega\,
	\textrm{d}\omega'\;.
\end{equation}
Substitution of Eqs.~(\ref{eq:Gamma_def}) and (\ref{eq:2D-PSD_inv}) into 
Eq.~(\ref{eq:O_jk.expl}) gives
\begin{eqnarray} \label{eq:O_jk.expand}
O_{jk}
&=&\frac{1}{(2\pi)^2}\frac{\kappa^2}{N^2}\sum_{p,q=1}^4
	\partial_\alpha S_p\,
	\partial_\beta S_q\,
	\sum_{l,l'=0}^{N-1}
	\int\!\!\!\int_{-\infty}^{+\infty}
	\textrm{d}\omega\,
	\textrm{d}\omega'\;
	P_{\alpha\beta}(\omega,\omega') \nonumber \\
&&{}\times
	\int_{-\infty}^{+\infty}
	Z_{pj}(t)\,
	\mathrm{e}^{\mathrm{i}\omega(t+t_{j,l})}\;
	\textrm{d}t 
	\int_{-\infty}^{+\infty}
	Z_{qk}(t')\,
	\mathrm{e}^{-\mathrm{i}\omega'(t'+t_{k,l'})}\;
	\textrm{d}t' \nonumber \\
&=&\frac{\kappa^2}{(2\pi)^2}\sum_{p,q=1}^4
	\partial_\alpha S_p\,
	\partial_\beta S_q
	\int\!\!\!\int_{-\infty}^{+\infty}
	\textrm{d}\omega\,
	\textrm{d}\omega'\;
	P_{\alpha\beta}(\omega,\omega')\,
	\tilde{Z}_{pj}^\ast(\omega)\,
	\tilde{Z}_{qk}(\omega') \nonumber \\
&&{}\times
	\frac{1}{N}\sum_{l=0}^{N-1}
	\mathrm{e}^{\mathrm{i}\omega t_{j,l}}\;
	\frac{1}{N}\sum_{l'=0}^{N-1}
	\mathrm{e}^{-\mathrm{i}\omega't_{k,l'}}\;.
\end{eqnarray}

The summations over the $N$ cycles in the last line can be performed
algebraically, using the rule for geometric series. If we recall 
Eq.~(\ref{eq:time}), and we introduce the total integration time $T=N n
\Delta t/r$ of the full polarimetric sequence, we find, e.g.,
\begin{eqnarray} \label{eq:sumrule}
\frac{1}{N}\sum_{l=0}^{N-1}
	\mathrm{e}^{\mathrm{i}\omega t_{j,l}}
&=&
\exp\biggl[ \mathrm{i} \omega\,(j-1)\frac{\Delta t}{r} \biggr]
\exp( \mathrm{i}\omega\tau)\,
\frac{1}{N}\sum_{l=0}^{N-1}
\exp\biggl( \mathrm{i} \omega\,ln\frac{\Delta t}{r} \biggr) \nonumber \\
&=&
\exp\biggl[ \mathrm{i} \omega\,(j-1)\frac{T}{Nn} \biggr]
	\exp\biggl[ \mathrm{i} \omega
	\biggl(\frac{N-1}{2N}\,T+\tau\biggr)\biggr]\,
\frac{\mathrm{sinc}(\omega T/2)}{\mathrm{sinc}(\omega T/2N)}\;. 
\end{eqnarray}
The second summation in Eq.~(\ref{eq:O_jk.expand}) is immediately
obtained via complex conjugation of Eq.~(\ref{eq:sumrule}), along 
with the substitutions $\omega\to\omega'$ and $j\to k$.
Thus, we finally obtain
\begin{eqnarray} \label{eq:O_jk.last}
O_{jk}
&=&\frac{\kappa^2}{(2\pi)^2}\sum_{p,q=1}^4
	\partial_\alpha S_p\,
	\partial_\beta S_q
	\int\!\!\!\int_{-\infty}^{+\infty}
	\textrm{d}\omega\,
	\textrm{d}\omega'\;
	P_{\alpha\beta}(\omega,\omega')\,
	\tilde{Z}_{pj}^\ast(\omega)\,
	\tilde{Z}_{qk}(\omega') \nonumber \\
&&{}\times
\exp\biggl[ \mathrm{i}\bigl[\omega\,(j-1)-\omega'(k-1)\bigr]\frac{T}{Nn} \biggr]
	\exp\biggl[ \mathrm{i} (\omega-\omega')
	\biggl(\frac{N-1}{2N}\,T+\tau\biggr)\biggr] \nonumber \\
&&{}\times
\frac{\mathrm{sinc}(\omega T/2)}{\mathrm{sinc}(\omega T/2N)}\,
\frac{\mathrm{sinc}(\omega'T/2)}{\mathrm{sinc}(\omega'T/2N)}\;. 
\end{eqnarray}

In Paper~I, the Fourier transforms $\tilde{Z}_{ij}(\omega)$ of the
functions (\ref{eq:Zcomp})
were calculated for both stepped and continuously rotating 
modulators (respectively, in Sect.~6 and App.~B). In particular, for a
stepped modulation scheme, we simply have
\begin{eqnarray}
&&m_i(t+t_{j,0})=m_{ji}\;, \\
&&\tilde{Z}_{ij}(\omega)=m_{ji}\,
	\Delta t\,\mathrm{sinc}(\omega\Delta t/2)\;.
\end{eqnarray}
In such a case,
\begin{eqnarray} \label{eq:O_jk.step}
(O_{jk})_{\rm step}
&=&\frac{\kappa^2\Delta t^2}{(2\pi)^2}\sum_{p,q=1}^4
	m_{jp} m_{kq}\,
	\partial_\alpha S_p\,
	\partial_\beta S_q
	\int\!\!\!\int_{-\infty}^{+\infty}
	\textrm{d}\omega\,
	\textrm{d}\omega'\;
	P_{\alpha\beta}(\omega,\omega') \nonumber \\
&&{}\times
\exp\biggl[ \mathrm{i}\bigl[\omega\,(j-1)-\omega'(k-1)\bigr]\frac{T}{Nn} \biggr]
	\exp\biggl[ \mathrm{i} (\omega-\omega')
	\biggl(\frac{N-1}{2N}\,T+\tau\biggr)\biggr] \nonumber \\
&&{}\times
\mathrm{sinc}(\omega \Delta t/2)\,
\mathrm{sinc}(\omega'\Delta t/2)\,
\frac{\mathrm{sinc}(\omega T/2)}{\mathrm{sinc}(\omega T/2N)}\,
\frac{\mathrm{sinc}(\omega'T/2)}{\mathrm{sinc}(\omega'T/2N)}\;. 
\end{eqnarray}

As a final verification, Eq.~(\ref{eq:sigmaN}) implies that the
matrix $\tens{O}_n$ must be purely real. To prove this, we first note 
that the modulation functions $Z_{ij}(t)$ are purely real, and thus
their Fourier transforms have the conjugation property 
$\tilde f^\ast(\omega)=\tilde f(-\omega)$. Similarly, from the
definition (\ref{eq:2D-PSD}), the jitter PSD satisfies the conjugation 
relation 
$P_{\alpha\beta}^\ast(\omega,\omega')=P_{\alpha\beta}(-\omega,-\omega')$.
The reality of the element $O_{jk}$ is thus
immediately demonstrated by performing the substitution
$(\omega,\omega')\to(-\omega,-\omega')$ in the double integral over 
$(\omega,\omega')$ in Eq.~(\ref{eq:O_jk.last}), which 
leaves unchanged the sign of said integral.

To summarize, Eq.~(\ref{eq:sigmaN}), where 
$\tens{D}$ is the Moore-Penrose inverse of the average modulation 
matrix $\tens{M}$ (see Eq.~(\ref{eq:mod_mat})), and 
the elements of $\tens{O}_n=(O_{jk})$ are given by
Eq.~(\ref{eq:O_jk.last}) (or Eq.~(\ref{eq:O_jk.step}), for 
a stepped modulator), provides the desired extension of the formalism of 
Paper~I to the estimation of polarization cross-talk produced 
in the presence of random image jitter that does not
satisfy the conditions of stationarity and ergodicity.

\begin{figure}[!t]
\centering
\includegraphics[width=.49\linewidth]{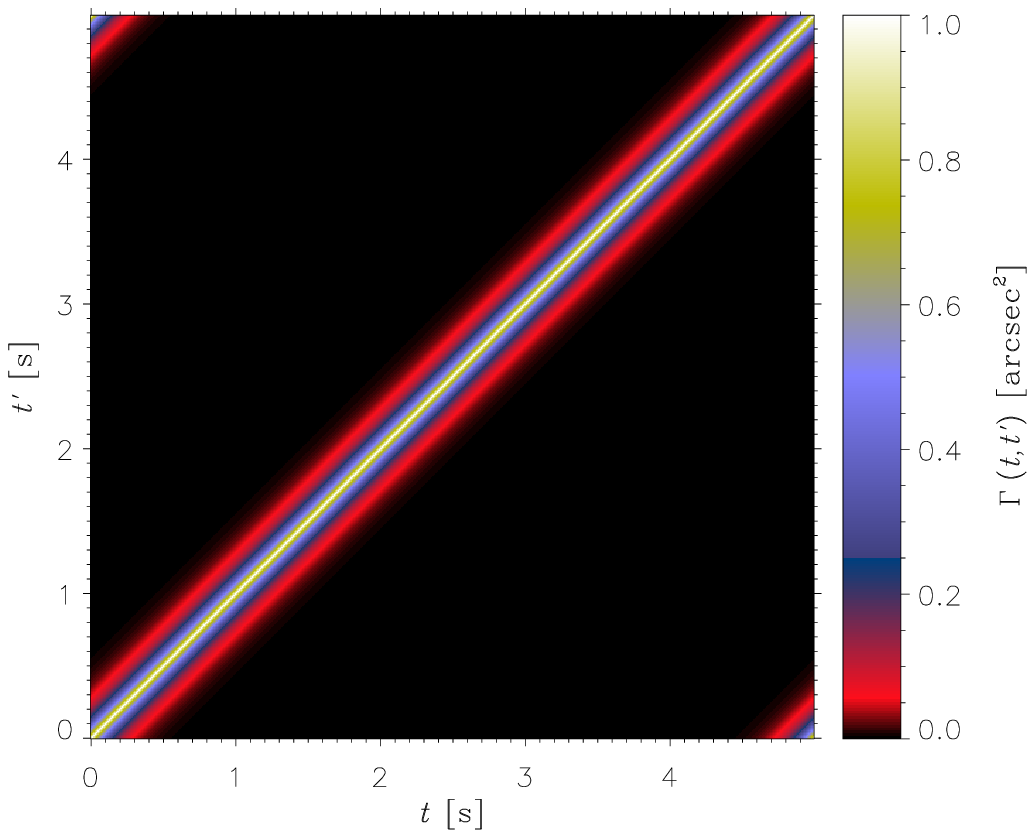}
\includegraphics[width=.49\linewidth]{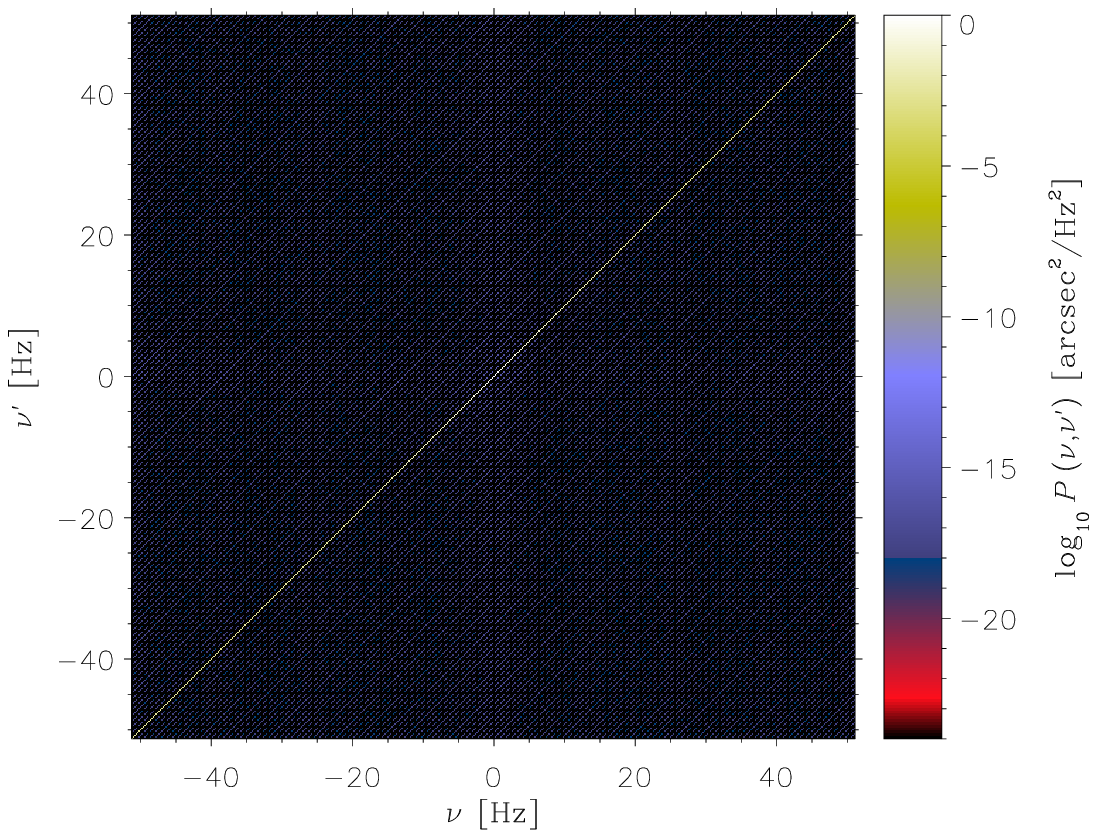}
\includegraphics[width=.49\linewidth]{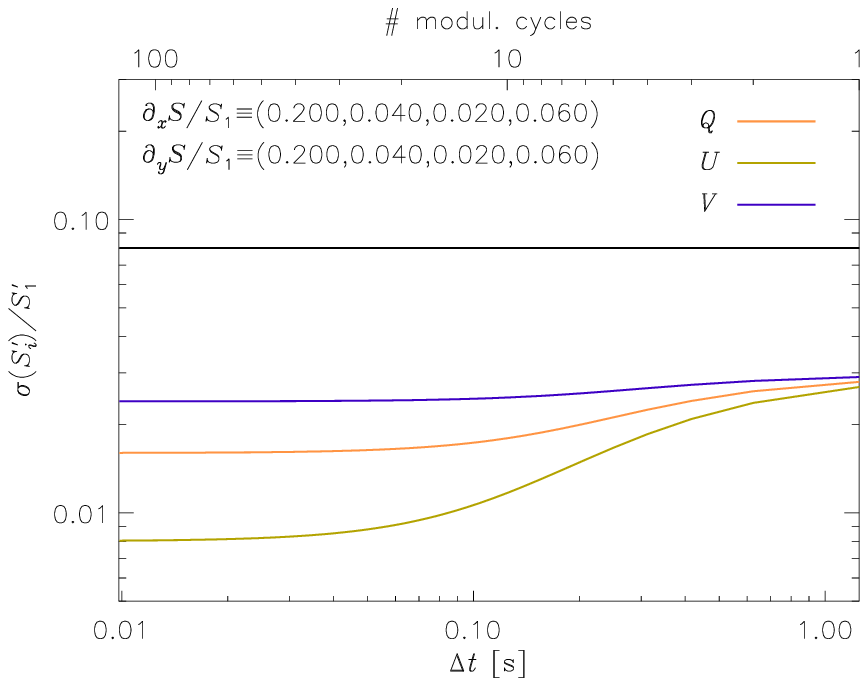}
\includegraphics[width=.49\linewidth]{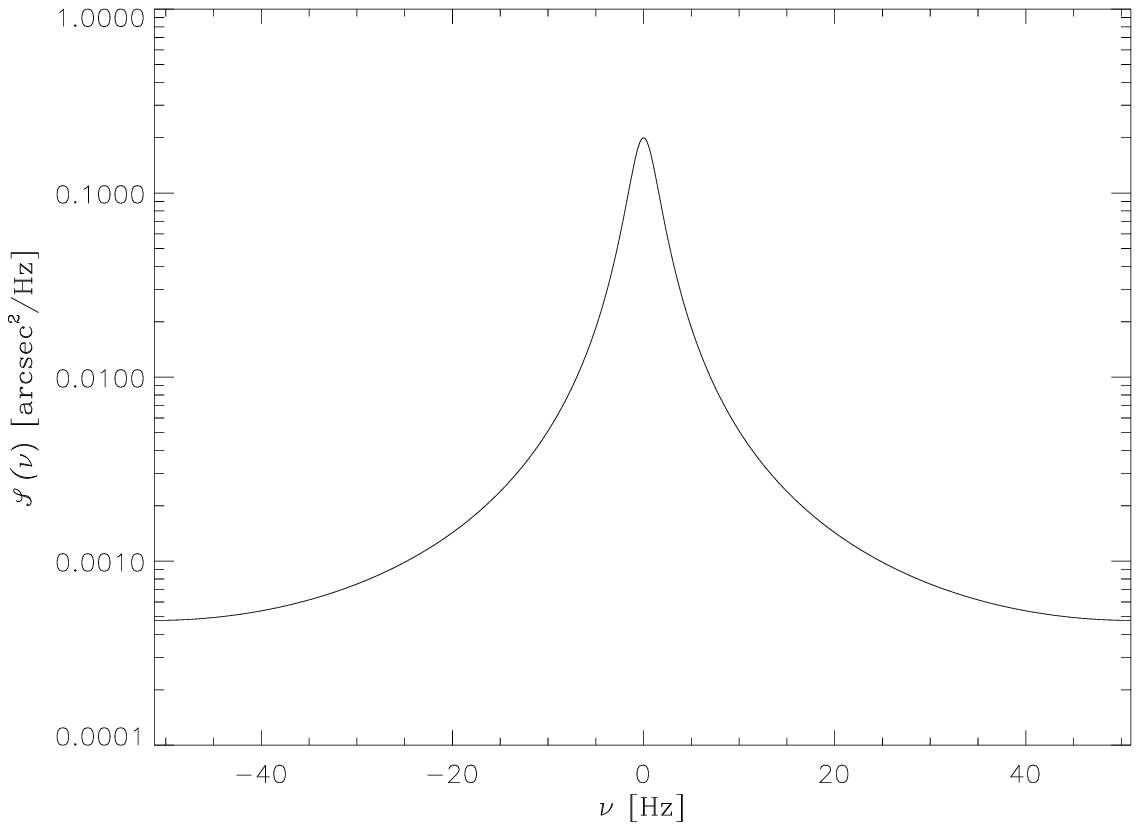}
\caption{\label{fig:circulant}
\emph{Top-left:} example of a two-time ACF of the \emph{circulant} form. 
\emph{Top-right:} the two-frequency PSD of the diagonal type 
$P(\omega,\omega')=2\pi\,\delta(\omega-\omega')\,\mathscr{S}(\omega)$ 
obtained as the Fourier transform of the two-time ACF. See text for details.
\emph{Bottom-right:} the 1D PSD corresponding to the diagonal of the 
2D PSD of the top-right panel. \emph{Bottom-left:} example of 
the Stokes cross-talk values induced by the image jitter 
associated with the 1D PSD, according to the model of
Eq.~(\ref{eq:O_jk.stat}), 
assuming typical Stokes gradients on the image plane (the gradient 
vectors are reported on the top-right of the panel); the cross-talk 
values are identical to those calculated using the full 2D PSD of
Fig.~\ref{fig:circulant}, using the model of Eq.~(\ref{eq:O_jk.last}).}
\end{figure}

\begin{figure}[t!]
\centering
\includegraphics[width=.495\linewidth]{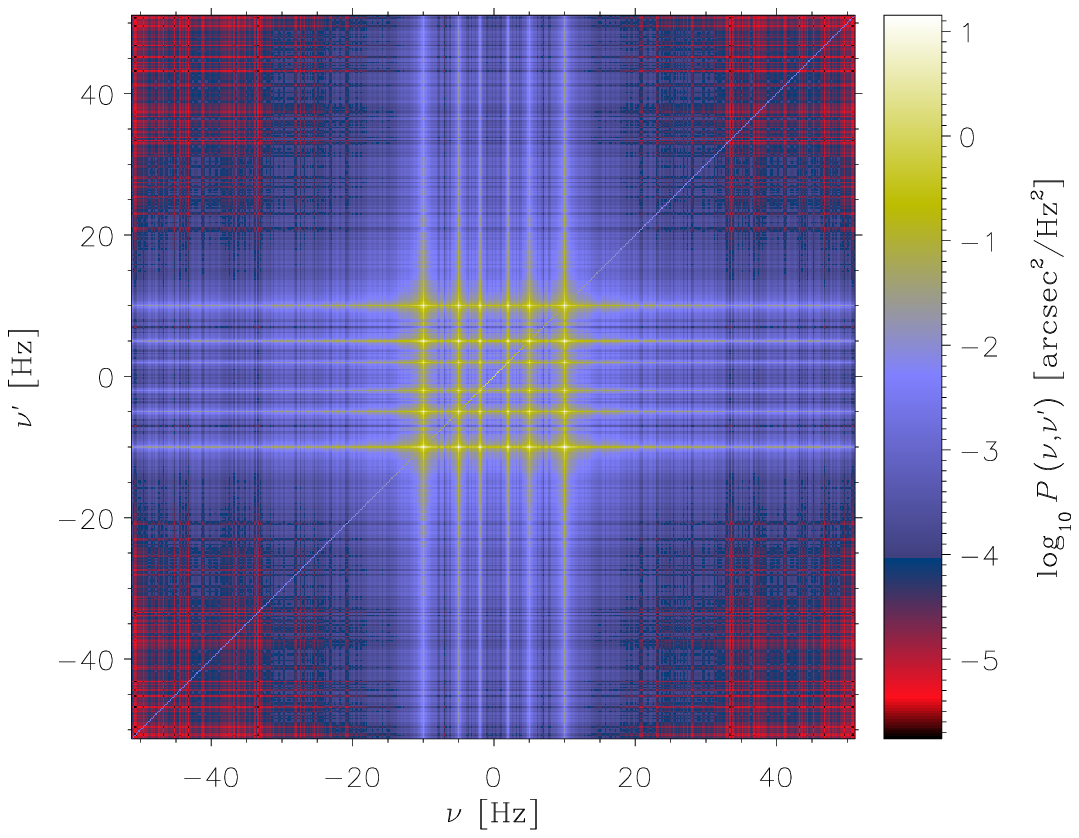}
\includegraphics[width=.495\linewidth]{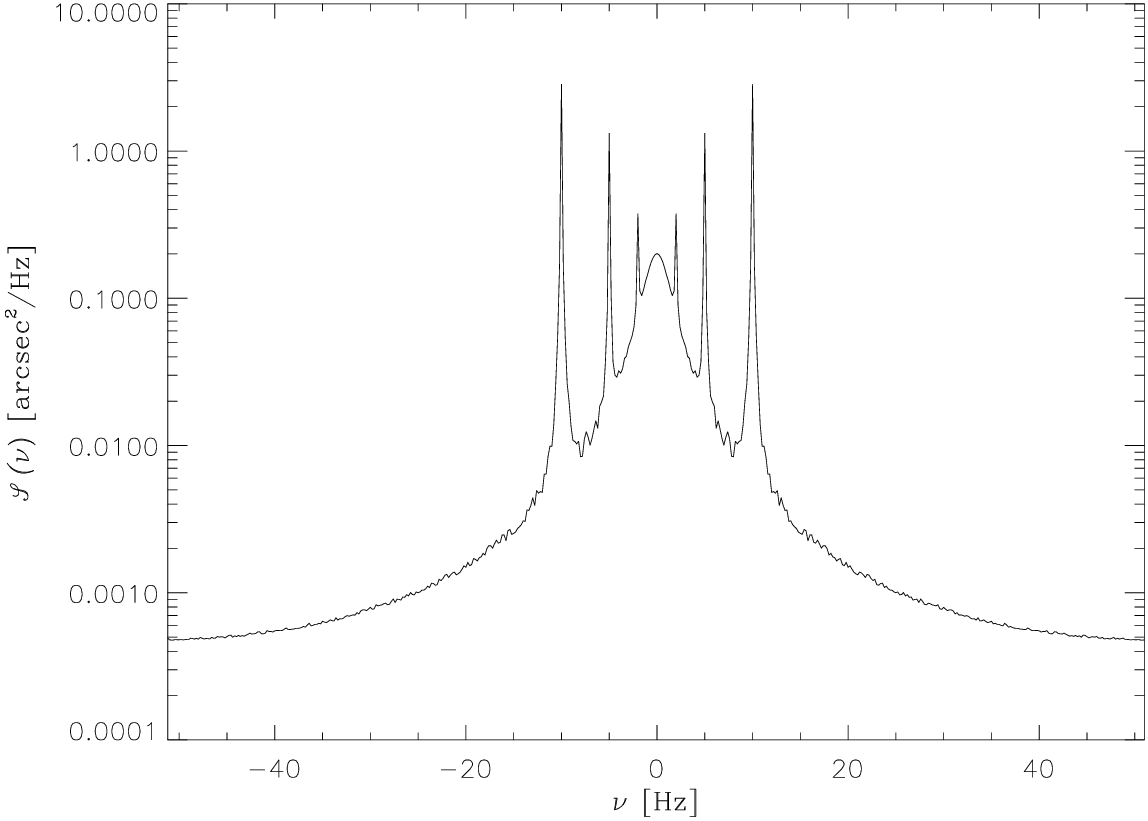}
\includegraphics[width=.495\linewidth]{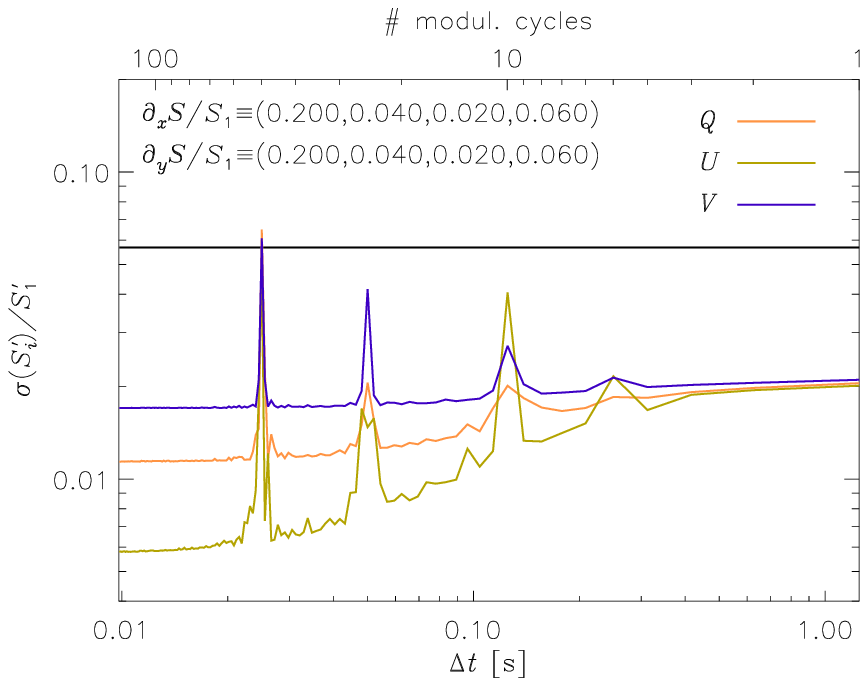}
\includegraphics[width=.495\linewidth]{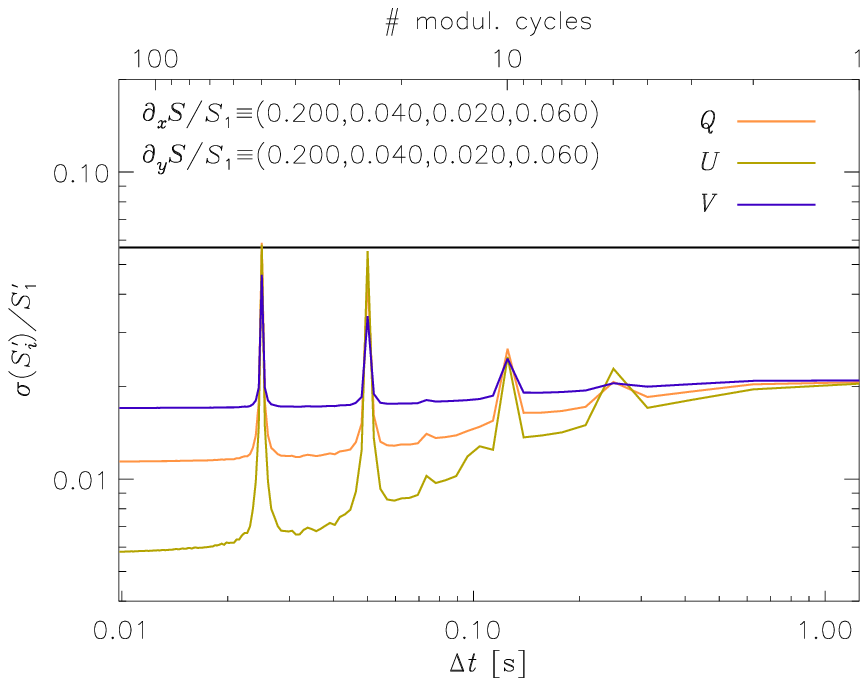}
\caption{\label{fig:NS}
\emph{Top-left:} the 2D PSD corresponding to the circulant 2D ACF 
of Fig.~\ref{fig:circulant}, after adding narrow power structures of 
different amplitudes located at 2, 5, and 10\,Hz. \emph{Top-right:} 
the 1D PSD corresponding to the major diagonal of the 2D PSD.
The bottom panels show the estimation of the polarization cross-talk
induced by the non-stationary random image jitter, using the same Stokes
gradients as in Fig.~\ref{fig:circulant}, using respectively the full 2D
frequency model of Eq.~(\ref{eq:O_jk.last}) (left), and the 1D frequency
model of Eq.~(\ref{eq:O_jk.stat}) applied to the 1D PSD shown in the
figure (right). We clearly see how the two calculations diverge exactly
in correspondence of the power spikes that violate the stationarity of
the underlying circulant ACF, because of the contribution to the
cross-talk of the off-diagonal power clearly manifested in the 2D PSD
map of the figure.} 
\end{figure}

\begin{figure}[t!]
\centering
\includegraphics[width=.495\linewidth]{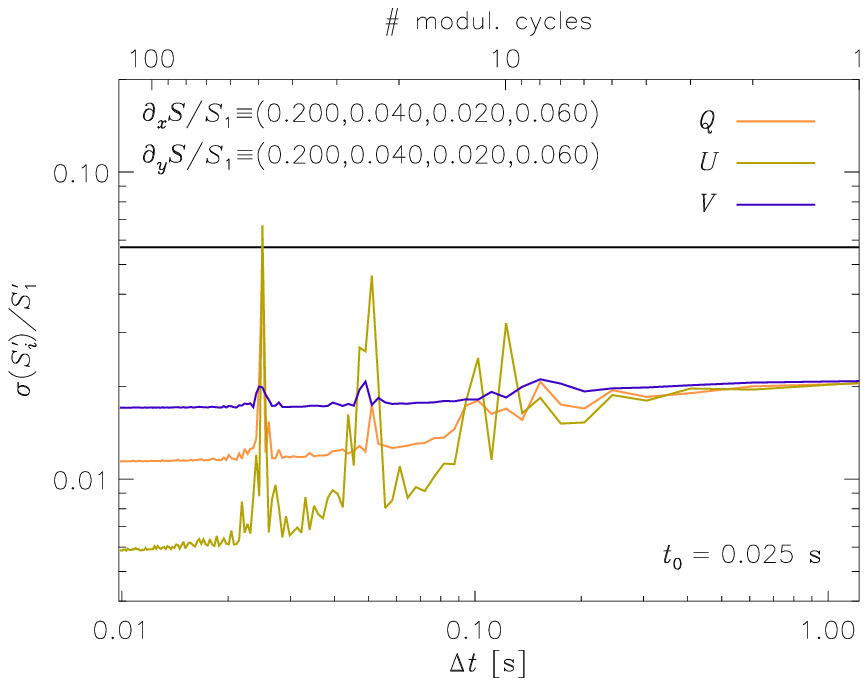}
\includegraphics[width=.495\linewidth]{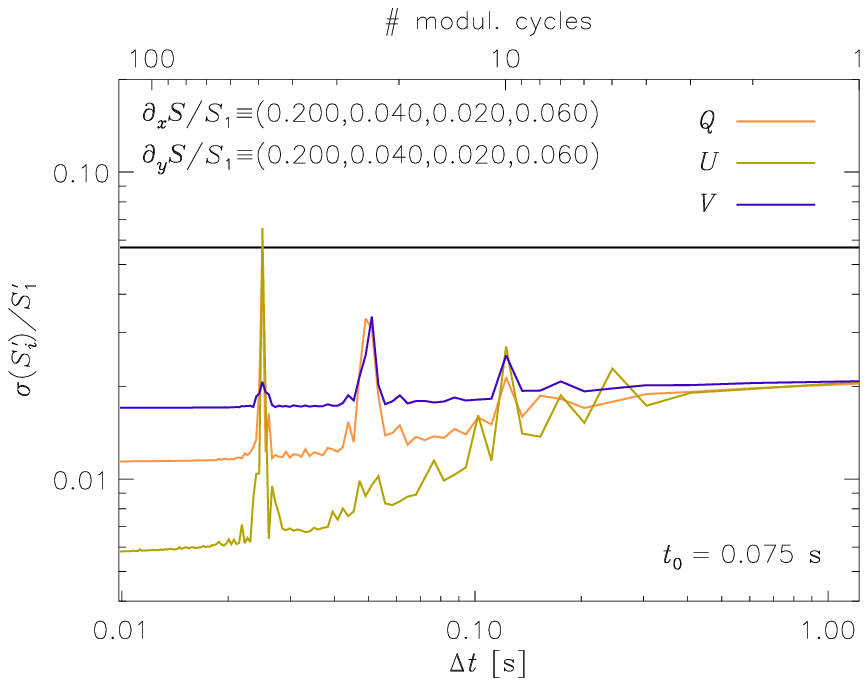}
\caption{\label{fig:NS_t0}
Polarization cross-talk plots corresponding to the non-stationary two-time ACF of Fig.~\ref{fig:NS}, but computed for two different starting times of the modulation sequence, $t_0=25$\,ms (left) and $t_0=75$\,ms (right),  from the beginning of the jitter time series. We note how the amplitudes of the cross-talk ``resonance'' peaks depend on the starting time of the modulation sequence in the case of a non-stationary time series. For comparison, the analogous plot in the bottom-left of Fig.~\ref{fig:NS} corresponds to the case $t_0=0$.}
\end{figure}

\section{Numerical tests}

Here we want to apply the formalism developed in the previous section to the estimation of 
polarization cross-talk in the presence of image instabilities that need 
to be modeled as non-stationary random processes. 

Before doing that, however, it is important to remark that the 
formalism derived in the previous section must reduce to the one 
presented in Paper~I, in the limit of stationary random motions, 
such as atmospheric seeing. 
To show this, we recall that the expectation value of the dyadic product 
\begin{displaymath}
x_\alpha(t+t_{j,l})\,x_\beta(t'+t_{k,l'})\;,
\end{displaymath}
which arises in the product
$O_{jk}=E(\delta\mathscr{I}_j\delta\mathscr{I}_k)$,
in the case of stationary random processes, 
is associated with the \emph{timeshift-invariant} ACF
$\Gamma_{\alpha\beta}(t,t')=\Gamma_{\alpha\beta}(t'-t)$. 
Then, it can be shown that the 2D PSD $P_{\alpha\beta}(\omega,\omega')$, 
which is obtained from such a function
via Fourier transform,
vanishes identically except along the singularity line $\omega=\omega'$. This 
is equivalent to introducing a $2\pi\,\delta(\omega-\omega')$ in the double 
integral of $O_{jk}$ (cf.~Eq.~(12-34) in \citealt{Pa65}). 
Under this circumstance, Eq.~(\ref{eq:O_jk.last}) gives explicitly
\begin{eqnarray} \label{eq:O_jk.stat}
(O_{jk})_{\rm stat}
&=&\frac{\kappa^2}{2\pi}\sum_{p,q=1}^4
	\partial_\alpha S_p\,
	\partial_\beta S_q
	\int_{-\infty}^{+\infty}
	\textrm{d}\omega\;
	\mathscr{S}_{\alpha\beta}(\omega)\,
	\tilde{Z}_{pj}^\ast(\omega)\,
	\tilde{Z}_{qk}(\omega) \nonumber \\
&&{}\times
\exp\biggl[ \mathrm{i}\,\omega(j-k)\,\frac{T}{Nn}
\biggr]\,
\frac{\mathrm{sinc}^2(\omega T/2)}{\mathrm{sinc}^2(\omega T/2N)}\;,
\end{eqnarray}
where $\mathscr{S}_{\alpha\beta}(\omega)$ is the corresponding 1D PSD
obtained as the Fourier transform of $\Gamma_{\alpha\beta}(t)$.

Equation~(\ref{eq:O_jk.stat}) leads to results in agreement with those of 
Paper~I (see, in particular, Sect.~6), if in addition we assume 
that the two stationary random processes $x_{1,2}(t)$ over the 
(Cartesian) image plane have zero average, so that
$\Gamma_{\alpha\beta}(t,t')=\delta_{\alpha\beta}\,\Gamma(t,t')$
(see Paper~I, before Eq.~(27)).

In order to verify this, 
we created a 2D ACF $\Gamma(t,t')$ of the \emph{circulant} 
type (a periodic special case of Toeplitz matrices; see 
\citealt{Gr06}), which has an exactly diagonal 2D PSD, thus formally
reducing Eq.~(\ref{eq:O_jk.last}) into Eq.~(\ref{eq:O_jk.stat}). 
In Fig.~\ref{fig:circulant}, the top-left panel shows such a 2D ACF, 
while the top-right panel shows the corresponding 2D PSD in logarithmic 
units. The bottom-right panel shows the plot of the 1D PSD corresponding
to the major diagonal of the 2D PSD.
The ACF was realized as a Gauss-Markov random motion 
$\Gamma(t,t')=\Gamma_0\,\exp(-\chi|t-t'|)$ (cf.~Paper~I, Eq.~(29)), 
with a temporal period of 5\,s, and assuming a variance value 
$\Gamma_0=1\,\rm arcsec^2$ and $\chi=10\,\rm s^{-1}$.
Along with its circulant property, the corresponding ACF represents
an ideally stationary random process, as confirmed by the 
diagonality of its PSD.

From a numerical perspective, in the case of a circulant ACF discretely sampled over a finite 
temporal domain of length $L$, we must note that the dimensional reduction from the 
2D PSD $P_{\alpha\beta}(\omega,\omega')$ to the 1D PSD 
$\mathscr{S}_{\alpha\beta}(\omega)$, via integration of the
Dirac-$\delta$ of the frequency, actually implies a normalization by $L$ of the 1D PSD, i.e.,
$P_{\alpha\beta}(\omega_k,\omega_k)=L\,\mathscr{S}_{\alpha\beta}(\omega_k)$,
where $\omega_k$ are the frequencies corresponding to the discretely 
sampled time intervals in the $[0,L)$ domain.

The polarization cross-talk estimations using the model of 
Eq.~(\ref{eq:O_jk.last}) applied to the 2D PSD $P(\omega,\omega')$
of Fig.~\ref{fig:circulant} match those obtained via the 
model of Eq.~(\ref{eq:O_jk.stat}) applied to the 1D PSD represented by
the major diagonal of the 2D PSD, with a relative error of $\sim 10^{-8}$. Such numerically identical cross-talk estimates are plotted 
in the bottom-left panel of Fig.~\ref{fig:circulant}, as a function of
the exposure time of a 4-state optimally efficient and balanced
modulation scheme (cf.~Eq.~(35) of Paper~I).

Next, we modified the circulant 2D ACF by adding power features to 
the corresponding 2D PSD located at 2, 5, and 10\,Hz, attributing
different amplitudes to those features but the same ``linewidth''.
The modified 2D PSD and the diagonal 1D PSD are shown in the top panels
of Fig.~\ref{fig:NS}, respectively, at the left and the right.
The estimation of the polarization cross-talk provided by
Eqs.~(\ref{eq:O_jk.last}) and (\ref{eq:O_jk.stat}), in this case, no
longer agree, showing clear differences exactly corresponding to the
power features that violate the stationarity of the original circulant
2D ACF; see bottom panels of Fig.~\ref{fig:NS}. These differences are
evidently associated with the power contributed by these structures
inside off-diagonal regions of the 2D PSD map, which are therefore
completely missed by the 1D PSD realized along the major diagonal, which
is described by the formalism of Paper~I.
Nonetheless, one can optimistically conclude from such a preliminary analysis that the cross-talk estimations predicted by the formalism developed in Paper~I can still be used to properly identify the temporal correlations between the modulation cycle and time series realizations of the image jitter, as well as for an approximate quantification of the magnitude of those correlations. 

To clarify this last point, it is necessary to discuss in more detail about both the locations and the amplitudes 
of the cross-talk spikes seen in Fig.~\ref{fig:NS}. The locations of such ``resonances'' always
coincide between the two stationary and non-stationary models, as they are fundamentally set by 
the frequency dependence of the integral kernels in 
Eqs.~(\ref{eq:O_jk.last}) and (\ref{eq:O_jk.stat}), which is identical in the two models, and because of the $(\omega,\omega')$ symmetries of the 2D PSD even in the case of non-stationary processes (see top-left panel of Fig.~\ref{fig:NS}). As to the amplitudes, instead, we can easily deduce from Eq.~(\ref{eq:O_jk.expand}) that any offset $t_0$ of the modulation sequence within the time series of the jitter implies the appearance of an
additional phase factor 
$\exp[\mathrm{i}(\omega-\omega')t_0]$
in the integrand of Eq.~(\ref{eq:O_jk.last}). While this leaves the diagonal of the 2D PSD unchanged, on the other hand it affects the phases of the off-diagonal contributions to the
cross-talk. For this reason, the amplitudes of the cross-talk resonances 
predicted by the non-stationary model strongly depend on the placement of the 
modulation sequence within the temporal realization of the image jitter, possibly even falling below those estimated by the 1D model, whereas
the stationary model of Paper~I is insensitive to it. 

This is clearly illustrated by the two examples of Fig.~\ref{fig:NS_t0}, where delays of 25\,ms and 75\,ms in the starting time $t_0$ of the modulation sequence were added in the calculation of the polarization cross-talk for the non-stationary time series of Fig.~\ref{fig:NS}.
Remarkably, 
by averaging over a sufficiently large number of $t_0$ values uniformly distributed over the length $L$ of the time series, the off-diagonal contribution to cross-talk is suppressed by phase cancellation, and we directly recover Eq.~(\ref{eq:O_jk.stat}) in that limit. As a consequence, the polarization cross-talk plots after such an averaging process converge to those for the stationary case shown in the bottom-right panel of Fig.~\ref{fig:NS}.
In other words, the stationary model
of Paper~I can be interpreted as estimating the expected cross-talk 
for a modulation sequence started at a random time relative to a given 
jitter realization, while the non-stationary model estimates the actual
cross-talk for that realization and a specific placement of the modulation 
sequence within it. This is a more precise statement about the possibility suggested in the previous paragraph that the simpler stationary model can still be used for a qualitative assessment of the impact of power resonances in the jitter PSD even when the presence of such resonances clearly indicates a departure from stationarity of the jitter mechanisms.


\section{Conclusions}

We have generalized the formalism of Paper~I for the estimation of
polarimetric cross-talk induced by image jitter to the case of random image
motions that are not stationary, and possibly not isotropic. The mean-square
error on the demodulated Stokes parameters is still given by
Eq.~(\ref{eq:sigmaN}), but the matrix $O_n$ is now determined by the
two-frequency PSD $P_{\alpha\beta}(\omega,\omega')$ of the two-time ACF of
the jitter, through Eq.~(\ref{eq:O_jk.last}) (or Eq.~(\ref{eq:O_jk.step}),
for a stepped modulator). Contrary to the stationary case, characterized
by the shift invariance of the ACF $\Gamma(t'-t)$, in the case of
non-stationary jitter processes, the resulting error estimate is specific 
to the actual location of the polarimetric sequence within the
realization of the jitter time series.

In the limit of a stationary random process, the 2D PSD is non-vanishing only along the
diagonal, $P_{\alpha\beta}(\omega,\omega')=2\pi\,\delta(\omega-\omega')
\,\mathscr{S}_{\alpha\beta}(\omega)$, and Eq.~(\ref{eq:O_jk.last}) reduces
identically to Eq.~(\ref{eq:O_jk.stat}), recovering the results of Paper~I.
We verified this numerically using a circulant 2D ACF \citep{Gr06}, for 
which the 2D PSD is exactly diagonal: the cross-talk estimates obtained from 
the full 2D model and from the 1D model applied to the diagonal PSD agree within the typical numerical noise of
the calculations (Fig.~\ref{fig:circulant}).

When the image jitter contains power characteristics that violate 
stationarity, the 2D PSD may include significant off-diagonal
structure---a contribution to the cross-talk that is entirely missed 
by the 1D PSD model, even when the latter is properly computed from the
same ACF (Fig.~\ref{fig:NS}). The differences are largest when the
modulation frequency $(n\Delta t)^{-1}$, or one of its harmonics, falls on a
non-stationary spectral feature. In such conditions, the stationary model
of Paper~I can misrepresent the cross-talk at precisely the exposure times
that correlate with the frequency of those power features, which is
something to be mindful of when designing a particular
polarization modulator and the associated modulation scheme. Nonetheless, the simpler stationary model can still be useful to correctly identify the critical rates of the modulation cycle, as well as providing an  approximate estimation of their cross-talk effects.

An important practical conclusion of this study is that the 1D PSD, which is often provided as a complete quantitative description of image jitter, does not 
carry enough information to recover the possible non-stationary character of the associated random process. The 1D PSD only determines the time-averaged ACF, and
there exists an infinite set of two-time ACFs that share the same time average. Therefore, the application of the present formalism
fundamentally relies on the availability of the two-time ACF, or which is
equivalent, of the time series of the jitter from which 
$\Gamma_{\alpha\beta}(t,t')$ can be estimated. As an alternative, the
same information can be produced utilizing realistic physical models of 
the image jitter (e.g., from knowledge of a spacecraft pointing system
and attitude control), from which an ensemble of statistically
independent realizations, and hence $\Gamma_{\alpha\beta}(t,t')$, can 
be generated.


\begin{acknowledgements}
This material is based upon work supported by the National Center for 
Atmospheric Research, which is a major facility sponsored by the 
National Science Foundation under Cooperative Agreement No.~1852977.

The authors used Anthropic Claude (Opus 5.5)  to help create and debug the IDL code
implementing the model of Eq.~(\ref{eq:O_jk.last}), starting from a previously 
existing code by the authors for the modeling of polarization cross-talk in the presence of stationary random sources of image motion based on Eq.~(\ref{eq:O_jk.stat}). Validation of the new code was performed by verifying it reproduced the correct stationary limit in the case of two-time ACFs described by ``circulant'' matrices; the use of such matrices for this validation (as well as the reference to \citealt{Gr06}) was suggested by Claude. The entire narrative and formal derivation, all numerical results, and the conclusions of this study were the exclusive product of the authors. (This statement can be refined during peer-review.)
\end{acknowledgements}

\end{document}